# Reproducibility Needs Reshape Scientific Data Governance

Paul Meijer, Yousef Aggoune, Madeline Ambrose, Aldan Beaubien, James Harvey, Nicole Howard, Neelima Inala, Ed Johnson, Autumn Kelsey, Melissa Kinsey, Jessica Liang, Paul Mariz, Stark Pister, Sathya Subramanian, Vitalii Tereshchenko, Anne Vetto

Allen Institute for Immunology

## Abstract

Scientific data governance should prioritize maximizing the utility of data throughout the research lifecycle. Research software systems that enable analysis reproducibility inform data governance policies and assist administrators in setting clear guidelines for data reuse, data retention, and the management of scientific computing needs. Proactive analysis reproducibility and data governance are integral and interconnected components of research lifecycle management.

In the field of research data management (RDM), open science principles inform appropriate research data collection, storage and management practices (Borghi & Van Gulick, 2022; Higman & Pinfield, 2015; Neylon, 2017a, 2017b; Stalla-Bourdillon et al., 2021; Waithira et al., 2019). Best practices include proper data storage for sharing, clear documentation for data organization and retrieval, and transparent sharing of data, tools and analysis methods. To establish these practices, institute-wide open science policies must be adopted, necessitating expertise, the creation of supporting infrastructure and a climate conducive to policy adoption.

Open science should be a core element of the lifecycle of any scientific study, and research organizations should support, recognize and celebrate scientists' open science efforts. As we propose in (Meijer et al., 2024), providing a computing environment that enables analysis reproducibility should be an integral part of that open science effort. In the current paper we show how the design principles of a scientific data analysis framework that enables analysis reproducibility inform data governance policies and can help administrators set clear guidelines around data reuse, data retention and the management of scientific computing needs.

# Research Lifecycle Management as a Policy

RDM affects all stages of the research lifecycle, from data acquisition and curation, to formatting the data for effective analysis and interpretation, to continuing data stewardship after the research ends (Borghi & Van Gulick, 2022). In other words, research data and tools management is an integral part of the research lifecycle and should be treated as such by scientists, rather than regarded as administrative overhead tangential to the research effort.

Despite the crucial role RDM plays in the success of any research endeavor, good management practices are often absent from the scientific curriculum (Tenopir et al., 2016) and are learned informally from laboratory peers instead (Borghi & Van Gulick, 2018, 2021). This absence of standardized management approaches can not only hinder individual research efforts, but also create significant inconsistency across research teams (Borghi & Van Gulick, 2022), unnecessarily complicating data reuse and sharing efforts.

Research organizations play a central role in defining the RDM approaches that govern the research lifecycle, supplying the know-how and infrastructure to support these approaches, and monitoring and rewarding their adoption (Janssen et al., 2020). Collaborations and open science initiatives can help standardize approaches across academic institutions.

# Analysis Reproducibility

In the past decade, a surge in recognition of the value of data management and sharing has been driven in part by concerns about the reproducibility of published studies. (Meijer et al., 2024) advocate for scientific computing frameworks designed with analysis reproducibility in mind. In their study, the authors present design principles that prioritize reproducibility and present a reference implementation of a reproducibility framework. This framework empowers scientists to proactively generate a real-time record of their research process. This trace is readily accessible for research team members' review before publication. On publication, scientists can share the data, methods and executable tools with their scientific findings.

By enabling proactive provenance tracking, this approach facilitates the early detection of unwarranted assumptions, oversights and errors, ideally during pre-publication team reviews. Furthermore, by providing clear insight into the study's entire methodology, the framework fosters a deeper understanding of the scientists' analytical choices. This transparency encourages discussion of alternative approaches within the scientific community.

As good stewards of scientific expertise, infrastructure and policy, research organizations have a pivotal responsibility to establish and promote the adoption of an analysis reproducibility framework. Next we explore how the foundational principles of the analysis reproducibility framework introduced by (Meijer et al., 2024) – proactive transparency, tracking data and transformation, advertising transparency, executable tooling, streamlining of administrative overhead through automation, and open science with equitable access – influence and guide RDM policy.

## Data Acquisition and Curation

To ensure that research methods and findings are reproducible, researchers should track their daily activities during the course of their work on a research project (Borghi & Van Gulick, 2022). The first guiding principle of analysis reproducibility introduced in (Meijer et al., 2024) is "Proactive Transparency." According to this principle, analysis frameworks should help researchers establish a ledger by automatically tracking all analysis activities in real time as the work progresses. This means that, for example, a data set is appropriately tagged as it's uploaded into the analysis tool. As data is filtered or curated, or a data set is generated as the output of an analysis step, each action produces a new data set that is similarly tagged. Specifically, as the second guiding principle, "Tracking Data and Transformation," expresses, both the data set and the processing of that data - for example, filtering, data normalization, quality control (QC), algorithmic analysis, or statistical derivation procedures - should be tracked to establish full provenance.

Combining these two principles, the full provenance can be depicted as a graph representing all data and transformations. Figure 1 shows a graph detailing the steps of an analysis culminating in a data visualization. In this example, multiple data files were uploaded into the analysis environment, and the file properties were recorded. Next a QC process validated these data sets to produce a single new data set. This data was used in an algorithmic analysis, and an analyst used its output in a coding environment to produce a visualization of the data. Note that each data set is immutable, and each analysis step is modularly tracked.

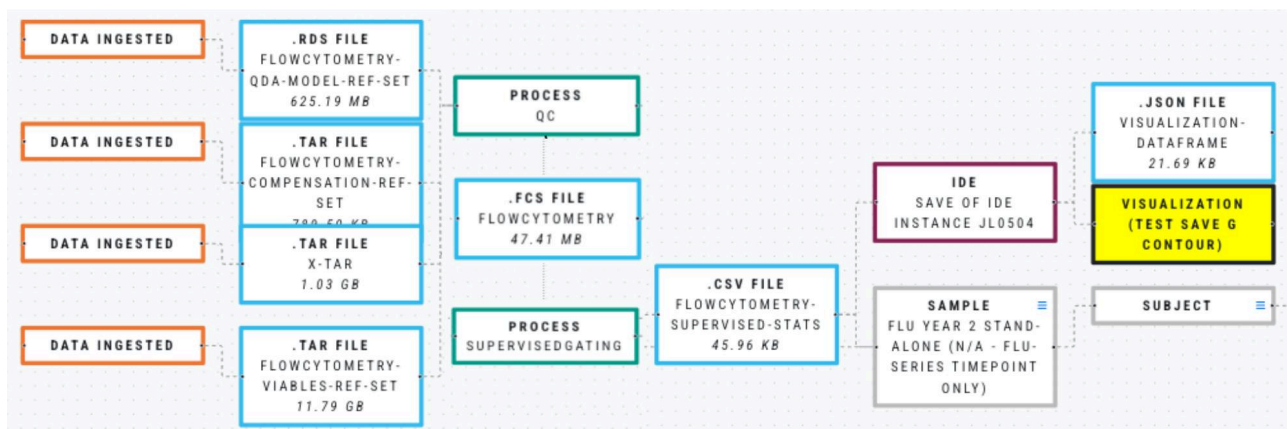

Figure 1. The full provenance of an analysis.

Effective RDM, then, must go beyond data management to include management of the tools and transformations used to evaluate data.

## Data Reuse

Effective RDM, exemplified by this proactive provenance tracking approach, unlocks the ability to re-use data throughout the active research phase. This meticulous tracking of every analysis step and intermediate data set allows researchers to readily share common root data (e.g., post–QC data) or apply the same analysis steps across different input data sets. Data reuse in turn encourages a collaborative team science approach throughout the research lifecycle and stimulates ongoing peer review before publication.

One of the core tenets of open science is that data is released at the end of a research study in order to foster reuse with the larger scientific community. Despite the recent emphasis on open science practices, released data assets still might not be suitable for reuse because of partial unavailability of (meta)data or unclear explanations of the origin of data (Roche et al., 2015). Efforts to provide analysis reproducibility to the scientific community significantly increase the potential for data reuse, especially when care is taken to include all relevant details of the analysis (Hardwicke et al., 2018, 2021). To that end, the entire provenance of the analysis should be shared in such a manner that data sets can be accessed and transformations examined and re-executed.

To reuse data, it must first be easily discoverable (Read et al., 2015). Our third guiding principle, "Advertising Transparency," underscores our belief that reproducible analyses should be clearly marked as such, for instance by publishing the analysis trace along with a prominently displayed marker or badge indicating that reproducibility principles have been followed. Being able to earn badges and actively display those badges to the community is effective in promoting open science practices (Kidwell et al., 2016). Additionally, it's essential to enable easy access to the analysis trace, for example by providing a digital object identifier (DOI) to facilitate direct citation of the trace and to designate the resource as a separate body of academic work.

## Data Transformation

Effective RDM practice goes beyond data management to include the management of tools and transformations. The next guiding principle, "Executable Tooling," captures the idea that transformations should be stored as instructions that can be re-executed at a later time. The set of criteria for what constitutes an executable instruction depends on the specifics of the underlying technology. Essentially, though, the instruction should produce a complete computational environment, including all analysis details and any necessary dependencies (Grüning et al., 2018; Heil et al., 2021; Perkel, 2023).

For instance, if an analyst runs one step of an analysis trace in a Jupyter notebook, the full original coding environment of that notebook is preserved, including the original (virtual) machine, third-party libraries, code, and instructions for accessing and loading the input data used in the analysis. In the cloud-based reference implementation discussed in (Meijer et al., 2024), any outside scientist may re-use this analysis setup in the same environment in which it was originally run, thus minimizing the possibility that environment drift could skew the outcome. Research teams should use open source technologies to minimize the likelihood of vendor lock-in and lessen the impact of product discontinuation (Bilder et al., 2015; Perkel, 2023). Additionally, open scholarly infrastructure and services require continual iteration (Lowenberg, 2022); doing so in the context of analysis reproducibility is warranted if system upgrades extend the lifetime of transformation availability without affecting reproducibility.

Over time, due to technological complications, such as third-party software libraries no longer being available, certain formerly executable instructions might no longer be executable. This inability to execute a transformation can have implications for either the data ingested into the transformation or the output data it generates. If the transformation can no longer be reproduced or the input data no longer serves as a source of reproducible analysis, the value of the output data must be reconsidered.

## Data Retention, Archival and Deletion

By prioritizing analysis reproducibility in data governance policies, data acquires a meaningful scientific context that transcends categorization based solely on IT criteria like age, size, file extension, or content type. For example, when raw data is transformed through a series of analysis steps to arrive at scientific insights, intermediate data sets might become candidates for data archival or even deletion provided those intermediate results can be easily regenerated. Similarly, if past executable transformations become unusable, the value of retaining the input and output data sets should be re-assessed. Provenance tracking endows data sets with semantic meaning based on their function within the analysis plan, enabling finer-grained data retention policies than IT criteria alone can provide.

Proactive transparency also offers valuable insights into fruitful data analysis approaches and abandoned paths. By tracking data access, policies can be devised to archive complete analysis traces – data and transformations – that are no longer being accessed and haven't been shared with the open science community.

## Policy Adoption

RDM policies can be effective only if researchers adopt them, and research organizations play a central role in encouraging and rewarding adoption. Supplying an analysis reproducibility framework provides tangential benefit to the research lifecycle, and its adoption promotes the RDM policies it reinforces.

Previous open science initiatives suggest that bureaucratic burden may hinder adoption and follow-through (Harris, 2018). Following the fifth guiding principle, "Streamlining Administrative Overhead through Automation," is a key condition of adoption, especially as all analyses are proactively tracked, whether they lead to publication or to inconclusive results. Automating as much of the provenance tracking process as possible is critical to lowering overhead. For instance, the preservation of an analysis done in a Jupyter notebook as an executable environment should be automated, and the analysts' actions should not be materially different from those made during untracked analysis. Lowering the threshold for adopting a provenance-compliant framework while rewarding scientists by highlighting their commitment to reproducible open science enables implicit adherence to the organization's RDM policies.

## Financial Sustainability

Data governance and open science policies must be financially viable (Bilder et al., 2015; Perkel, 2023; Waithira et al., 2019). Scientific analysis increasingly depends on substantial resources, including the necessary infrastructure to support computational and storage needs, as well as the expertise required to operate that infrastructure and to upgrade it as technologies evolve. Open access to scientific analysis goes beyond data sharing, requiring tools and resources for broader verification by the scientific community (Cole et al., 2022).

The primary self-correcting mechanism in science is the scientific community's active engagement in debate and evaluation of findings (Oreskes, 2021). To foster such discourse and effectively identify and rectify implicit assumptions, it's essential for published research to be accessible and interpretable by a diverse scientific community (Oreskes, 2021; Ross-Hellauer, 2022). The final guiding principle, "Open Science with Equitable Access," demands that any analysis reproducibility framework incorporate mechanisms for granting open access to data, transformations, infrastructure and the knowledge required to utilize them, thereby addressing the structural disparities that exist among scientists and academic institutions (Bezuidenhout & Chakauya, 2018). This guideline can be effective only if long-term financial health is safeguarded. If an institution incurs costs by sharing data and offering analysis tools to outside scientists, it must ensure that those costs are covered, for instance by charging usage fees or by setting aside part of their internal budget to subsidize open science projects.

Although analysis reproducibility and RDM policies may impose initial financial burdens on academic institutions, they ultimately contribute to better long-term financial planning for scholarly infrastructure and cost management. Gaining a precise understanding of internal and community-wide data and infrastructure utilization clarifies which components need maintenance and support. Conversely, it allows for the identification of unused resources, facilitating cost-effective decisions about archival, deletion, or discontinuation of services.

# Conclusion

Analysis reproducibility streamlines the implementation of data governance policies throughout the research lifecycle. To ensure sustained adoption, the scientific community must champion long-term funding for research infrastructure. Currently, the financial burden often falls on research institutions (Higman & Pinfield, 2015). Given that research outputs are expected to be accessible beyond grant durations, it's crucial to explore new funding models for sustainable infrastructure.

The rapidly evolving nature of software technology poses significant challenges to the long-term usability of research tools. A visit to any museum dedicated to computing history, such as the Computer History Museum or the National Museum of Computing, underscores the swift transformation of computing technology. These museums showcase the transition from early computing devices to mainframes, personal computers, and modern cloud-based systems, as well as the parallel development of software applications and computing capabilities (Computer History Museum, 2024; The National Museum of Computing, 2024). As we generate larger and more complex datasets and become increasingly reliant on advanced computing technology to extract insights, it's imperative to consider how this dependence impacts the long-term accessibility and utility of scientific data. As scientists, we have a responsibility to preserve our data, tools, and methods for the benefit of future researchers.

# Acknowledgement

We are thankful to all members of the Allen Institute for Immunology for their support for and dedicated contributions to the Human Immune System Explorer (HISE). We are indebted to Scott Pegg for the inspiring discussions about reproducibility and data governance, and are grateful to the leadership and support of Ananda Goldrath Executive Vice President, Director of the Allen Institute for Immunology, Rui Costa, President and CEO of the Allen Institute, and Allen Institute founder, Paul G. Allen, for his vision, encouragement, and support.